# Structural Damage Detection Using Ensemble Empirical Mode Decomposition, Hilbert Transform and Artificial Neural Networks

**Ali Bakhshi[1], Omid Bahar[2], and Sayyed Mohsen Vazirizade[3]***

Department of Civil Engineering, Sharif University of Technology, Tehran, Iran

1- Assistant Professor of Sharif University of Technology, Tehran, Iran - bakhshi@sharif.edu , Tel : +98 (21) 6601-3201
2- Associate Professor of International Institute of Earthquake Engineering and Seismology, Tehran, Iran - m-omidbahar@iiees.ac.ir , Tel : +98 (21) 2283-1116
3- M. Sc. Student of Sharif University of Technology, Tehran, Iran - corresponding author- s.m.vazirizade@gmail.com, - Phone:+989366927599

Abstract

Civil structures are on the verge of changing which leads energy dissipation capacity to decline. Structural Health Monitoring (SHM) as a process in order to implement a damage detection strategy and assess the condition of structure plays a key role in structural reliability. Earthquake is a recognized factor in variation of structures condition, inasmuch as inelastic behavior of a building subjected to design level earthquakes is plausible. In this study Hilbert Huang Transformation (HHT) is superseded by Ensemble Empirical Mode decomposition (EEMD) and Hilbert Transform (HT) together. Albeit analogous, EEMD brings more appropriate Intrinsic Mode Functions (IMFs) than Empirical Mode Decomposition (EMD). IMFs are employed to assess first mode frequency and mode shape. Afterward, Artificial Neural Networks (ANN) is applied to predict story acceleration based on acceleration of structure during previous moments. ANN functions precisely. Therefore, any congruency between predicted and measured acceleration provides onset of damage. Then another ANN method is applied to estimate stiffness matrix. Though first mode shape and frequency is calculated in advance, it essentially requires an inverse problem to be solved in order to find stiffness matrix. This task is done by ANN. In other words, these two ANN methods are exercised to forecast location and measure severity of damage respectively. This algorithm is implemented on one nonlinear moment-resisting steel frame and the results are acceptable.

1. Introduction

Civil structures in general and those subject to seismic excitation in particular are vulnerable to damage and deterioration during their service lives. Inasmuch as damages in civil systems have led to unwanted major loss and casualty, they have gained increasing attention from the scientific community. The process of assessing the condition of a structure in order to detect any imperfection is done by visual inspection traditionally and some other banal methods. These antediluvian methods are susceptible to be obsolescence. Thus, new methods, that is, Non-Destructive Evaluation techniques (NDE), entered into the arena. Although these such new methods as acoustic signals, electromagnetic, radiography, fiber optics and so forth are not only more effective and convenience but also more economical, these damage detection methods are not global, but local. Therefore, they are effective only for small structures or structural members[1]. Global vibration-based techniques have been released recently in order to overcome this challenge. These methods contain Fourier transform, power spectrum and spectrum analysis, to name but a few[2]. During the preceding decade, the use of vibration data to find an effective strategy for the purpose of quantifying structural damage in engineering structures has gained increasing attention from the scientific community. A review of vibration-based health monitoring methods can be found in [3–6]. The vibration-based methods, which only consider simply time are really worthwhile. Nonetheless, it should be mentioned that these methods discard time information. In order to solve this problem, time-frequency methods are implemented, which presents both frequency and time simultaneously. During the preceding decades, the use of vibration data to find an effective strategy for the purpose of quantifying structural damage in engineering structures has received considerable attention from the scientific community because of its ability to analyze the non-stationary of signals [7,8], and Hilbert Transform (HT) for signal processing is another kind of time-frequency analysis method that has been used for structural damage detection[9–11]. In other words, most of the signals, in real world, are not only nonlinear but also non-stationary concurrently which makes requires new methods rather than conventional ones. Wavelet Transform (WT) and Short Fast Furrier Transform (SFFT) are other tools for time-frequency analysis, and the former also provides variable-sized regions for windowing[12], by doing so, higher frequency resolutions is brought, and a uniform resolution for all scales is provided [13]. Apart from these methods, Huang et al. [14,15] firstly developed a new method—Hilbert Huang Transformation (HHT)—in order analyze both nonlinear and non-stationary signals. This aforementioned method is composed of two portions, that is, Empirical Mode Decomposition (EMD) and HT. In fact, the first step, so-called EMD, is a procedure in order to provide some signals, which stem from the main one, that admit well-behaved HT. In other words the mother signal is decomposed into Intrinsic Mode Function (IMF) since the EMD is based on the local characteristic time scale of the original data, this decomposition method is adaptive and highly efficient. The HHT method, in a galloping rate. has been employed in many scientific and engineering disciplines in general and structural health monitoring in particular to give new insights into the non-stationary and nonlinear signals [16]. Huang et al. [17] asserted that not only HHT was a more precise definition of particular events in time-frequency space than wavelet analysis, but this method was more physically meaningful interpretations of the underlying dynamic processes as well. Vincent *et al.* [18] compared the EMD method and wavelet analysis for structural damage detection. The structure, which is modeled as a simple 3 DOF system, is monitored during 20 seconds, the stiffness of first story decreases, and the instantaneous frequencies calculated from the first five IMFs. It is concluded that both the EMD and Wavelet methods are effective in detecting the damage. However, the EMD method seems to be more promising for quantifying the damage level. Many assorted methods are proposed to improve HHT both in decomposition part, EMD, and HT portion. In this regard Ensemble EMD (EMD) is presented utilizing the benefits of the properties of white noise to distribute component with more proper scales due to resolve the mode mixing problem in EMD [19,20].

In the very modern era, in which we live, improvements in computers' processing power pave the way for more elaborate computations. Artificial intelligence, therefore, has become the center of attention during the preceding decades. Artificial Neural Networks (ANNs) in terms of machine learning and cognitive science are learning models inspired by biological neural networks [21]. Suresh *et al.* [22] considered the flexural vibration in a cantilever beam. A neural network was trained by modal frequency parameters–calculated for various location and depth of crack. Furthermore, they investigated two widely used neural networks, namely the multi-layer perceptron network and the radial basis function network, and figured out the latter is better than first one by virtue of performance and less computational time. The effects of three different learning rate algorithms, the DSD algorithm, the FSD algorithm, and the TSD algorithm on the neural network training, are studied. Fang *et al.* [23] studied the back-propagation neural network (BPNN) and using frequency response functions (FRFs) as its input data in order to assess damage conditions on a cantilevered beam. Eventually, they asserted that this new approach was highly accurate in predicting damage location and severity and could be rewarding. Xu *et al.* [24] presented a method for the direct identification of structural parameters based on neural networks. They used two back-propagation neural networks the first of which, called emulator, was employed to forecast its dynamic response with sufficient accuracy using time-domain dynamic responses. The difference between structure response and neural network prediction assessed with root mean square (RMS). The other neural network - called parametric evaluation neural network - proposed to identify the structural parameters. Saadat *et al.* [25] proposed a method, that is the intelligent parameter varying (IPV), which uses radial basis function networks to estimate the constitutive characteristics of inelastic and hysteretic restoring forces and showed effectiveness of this method. Bandara *et al.* [14] utilize neural network for the actual damage localization and quantification based on frequency response function (FRF). Their network inputs were damage patterns of different damage cases which associate with FRF and outputs were either the locations of the damage or the damage severities. In this study, a methodology is proposed to detect damage in nonlinear moment-resisting steel frames. In fact, this method discern changes in structure stiffness and onset of damage and predict location and measure severity of damage by implementing EEMD, HT an ANNs. First, first mode frequency and mode shape are calculated thank to EEMD and HT process. Afterwards, by using two ANNs location and severity of damage is determined respectively.

2. Signal Processing Procedures

2.1. Hilbert Transform:

HT proves useful to compute instantaneous frequency, by doing so, complex conjugate $y(t)$ of any real valued function can be calculated. HT of a signal, $y(t)$ is defined by

$$H[x(t)] = y(t) = \frac{1}{\pi} PV \int_{-\infty}^{+\infty} \frac{x(t)}{t-\tau} d\tau$$

where $t$ is the time variable and PV indicates the principal value of the singular integral. With the HT, the analytic signal, $z(t)$, is obtained as

$$Z(t) = x(t) + iy = a(t)e^{i\theta(t)}$$

$$a(t) = \sqrt{x^2(t) + y^2(t)}$$

$$\theta(t) = \arctan\left(\frac{y}{x}\right)$$

in which $a(t)$ and $\theta(t)$ are the instantaneous amplitude and phase function respectively. Instantaneous frequency drives from derivative of phase function [26]:

$$f(t) = \frac{1}{2\pi} \frac{d\theta(t)}{dt}$$

In spite the fact that HT proves useful, through which to achieve physically meaningful instantaneous frequencies encountered some difficulties. There are some conditions and theories which express these shortcomings, for example Bedrosian and Nuttall theorems to name but a few. In this regard, many assorted methods such as Normalized Amplitude Hilbert Transform (NAHT) method and Enhanced Hilbert Huang Transform has been proposed [27,28]. Reducing the function into IMFs has improved the chance of getting a meaningful instantaneous frequency. Thus, EMD is procedure decompose a signal into its Intrinsic Mode Functions (IMFs)

2.2. Empirical Mode Decomposition (EMD):

In 1996, EMD was proposed by Huang *et al.* [15] for nonlinear and non-stationary data. As mentioned earlier, EMD is procedure decompose a signal into its IMFs an IMF is defined as a function that satisfies two conditions:

1. The number of extrema and the number of zero crossings must either equal or differ at most by one in the whole data
2. At any point, the mean value of the envelope defined by the local maxima and the envelope defined by the local minima is zero.

To this end, amplitude and frequency of IMFs can be non-constant and alters. In other words, amplitude and frequency are variable as a function of time, and in doing so, HT becomes a helpful transform.

For such a signal as $x(t)$ EMD procedure can be summarized in six steps as follows:

1. Determining all the local extrema, that is, maxima and minima.
2. A cubic spline passing all the local maxima as the upper envelope, $e_{max}(t)$, and an analogous procedure for the local minima to achieve the lower envelope, $e_{min}(t)$
3. Computing the mean value of upper envelope and lower one: $m_{i,j} =[ e_{min}(t)+ e_{max}(t)]/2$ where $i$ and $j$ indices indicates the number of associated IMF and iteration respectively. For first IMF, $i$ equals 1
4. Subtracting $m_{i,j}$ from the initial signal to find the first component, $h_{i,j}(t)$

$$h_{i,j} = x - m_{i,j}$$

5. Repeating steps 1 to 3 on $h_{i,j}(t)$ $k$ times to obtain the next IMF, $C_i(t)$

$$h_{i,k} = h_{i,j-1} - m_{i,j}$$

$$C_i = h_{i,k}$$

6. Calculating $r$ and repeating steps 1 to 5 to extract remaining IMFs

$$r_{i+1} = x - C_i$$

and a residue, $r_{n+1}(t)$,

Finally, by using these steps, $x(t)$ is decomposed to $n$-separate IMFs and a residue, $r_{n+1}(t)$. The original signal can be reconstructed by summation of intrinsic modes and the residue:

$$x = \sum_{i=1}^{n} C_i + r_{n+1}$$

Among the IMFs, high-frequency content is removed gradually from $C_1(t)$ to $C_n(t)$, resulting in the latter signals with lower and lower frequency content.

### 2.3. Ensemble Empirical Mode Decomposition (EEMD):

Wu and Huang presented EEMD [19,20] so as to solve the mode mixing problem by adding a white noise to the original signal. Mode mixing, which is consequent of signal intermittency, causes existence of disparate scales in an IMF or presence of a similar scale in different IMFs. Consequently, this drawback makes IMFs physically meaningless and it is crucial to get rid of it. Therefore, EEMD is implemented for not only surmounting mode mixing problem but also reduce the sensitivity of a signal to noise pollution, which is shown in this study. The statistical characteristics of white noise in company with an ensemble of trials bring about improvement in the scale separation problem. To put it differently, adding noise besides EMD and iteration afford better sifting process. It should be mentioned that though the IMFs are polluted by noise in each trial, by utilizing ensemble mean, this effects are cancelled out. Wang *et al.* [29]compared the applications of EMD and EEMD on time-frequency analysis of seismic signal. They found out that this new approach is remarkably capable to solve the mode mixing problem. They demonstrated this assertion by applying an example and revealed EEMD ability to decompose the signal into different IMFs and analyze the time-frequency distribution of the seismic signal.

EEMD is summarized by the following steps:

1- Adding white noise $w$ to the initial signal, $x(t)$, for the first try $k=1$ and for the first IMF $r=1$ as well:

$$x_1 = x + w_{1,k}$$

And for other step the formula is presented as:

$$r_{i,k} = r_i + w_{i,k}$$

2- Applying steps 1 to 5 of EMD procedure to decompose the signal to its IMF
3- Repeating steps (1) to (2) with different white noise for each try
4- Averaging corresponding IMF calculated by each trial in order to obtain $C$

$$C_i = \lim_{n \to \infty} \frac{1}{n} \sum_{k=1}^{n} C_{i,k}$$

Where n is number of ensemble

5- subtracting $C_i(t)$ from the initial signal and repeat these steps in order to find remaining IMFs

And finally, the main signal can be reproduced as following equation:

$$x = \sum_{i=1}^{n} C_i + r_{n+1}$$

The advantage of EEMD over EMD will be shown in numerical example in the following; however a simple example verifies the fact that EEMD is less sensitive to noise than EMD.

The following equation demonstrate the signal, y, which contain three assorted frequencies, 2, 4, 8 and the exponential part simulate the attenuation.

$$y = \left[5\,cos(2\pi \cdot 2t) + 5cos(2\pi \cdot 4t) + 2sin(2\pi \cdot 8t)\right]e^{-t/10}$$

This signal is polluted with various levels of noise as illustrated in Figure 1, where signals $y_1$, $y_2$, $y_3$, and $y_4$ polluted with different noise levels 0, 5%, 10%, and 20% respectively. In this regard, Figure 2 and Figure 4 shows IMFs for mentioned signals extracted by EMD and EMD respectively, and Figure 3 and Figure 5 presents their corresponding HT. It is crystal clear that EMD is so sensitive to noise level and its effect on IMFs is conspicuous. On the other hand, the IMFs, which extracted by EEMD, are more analogues with each other.

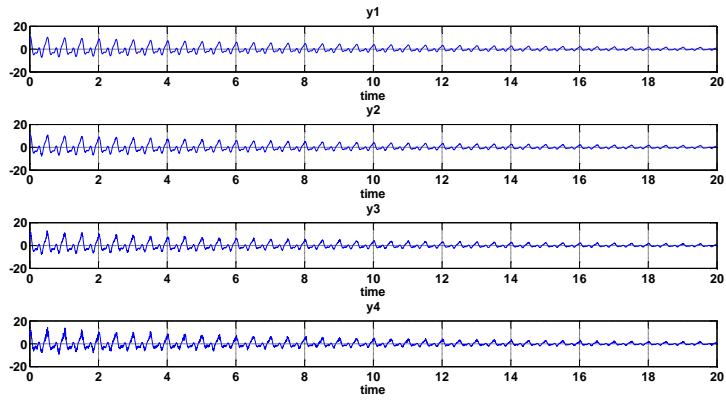

*Figure 1 signal y which is polluted with different noise level*

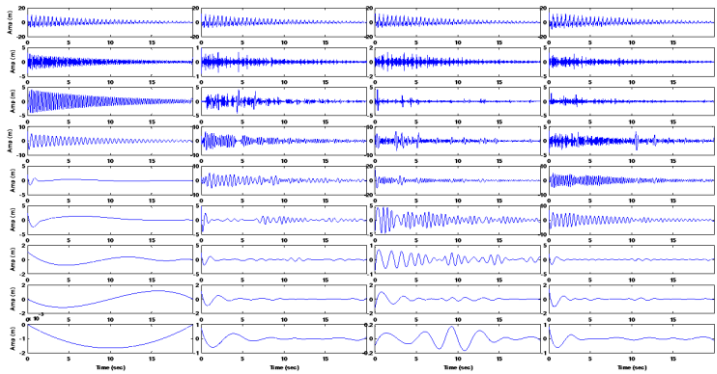

*Figure 2 IMFs for four different noise levels extracted by EMD*

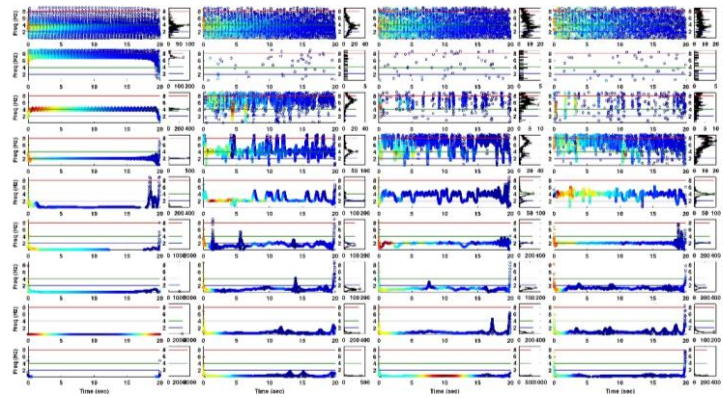

*Figure 3 Applying HT to IMFs extracted by EMD*

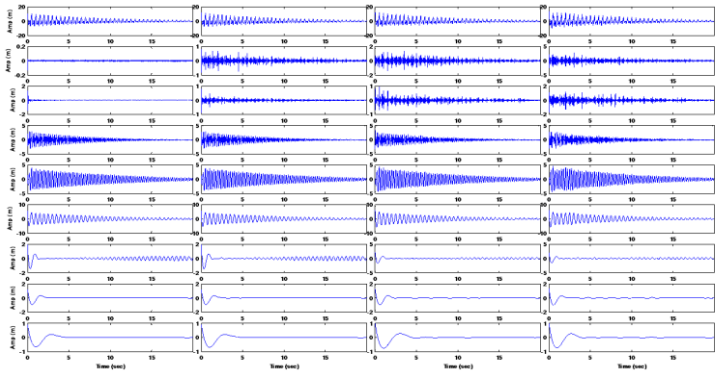

*Figure 4 IMFs for four different noise levels extracted by EEMD*

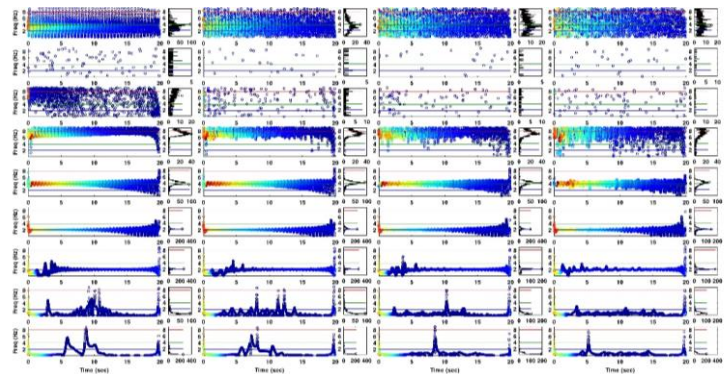

*Figure 5 Applying HT to IMFs extracted by EEMD*

In fact, this uniformity among decomposed signals is of considerable importance. Normalized IMFs with similar frequency content relative their corresponding IMF related to $y_4$ are presented in Figure 6 for EMD and Figure 7 for EEMD. As it is shown because of non-uniformity among the IMFs, normalized amplitudes undergo more anomalies in Figure 6 though all the $y_1$ to $y_4$ signals stem from $y$ and simply polluted with different noise level and due to this fact their IMFs should be almost the same.

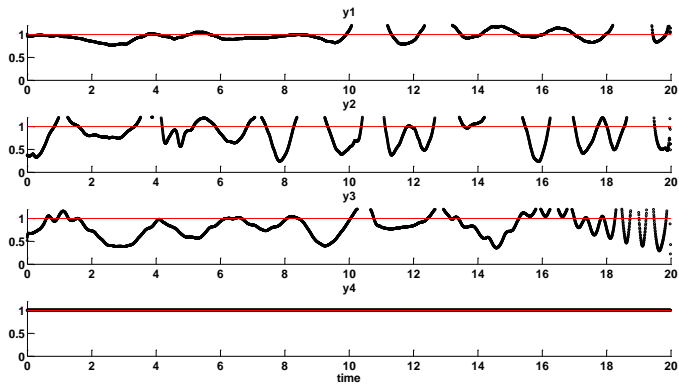

*Figure 6 Normalized IMFs number4, 5, 6, 6 for $y_1$, $y_2$, $y_3$, and $y_4$ respectively relative to number 6 for $y_4$*

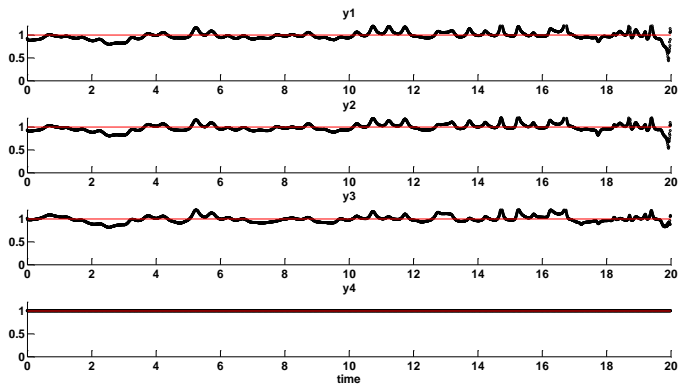

*Figure 7 Normalized IMFs number 6 for y1 to y4 relative to number 6 for y4*

3. Damage detection procedure

In this section, the rewarding process of finding severity and location of damage, which is a combination of signal processing and artificial intelligence is proposed. Figure 8 plots the schematic process of proposed method for a typical 3-story moment-resisting frame. In the following steps, we will delve into the procedure scrutinizingly.

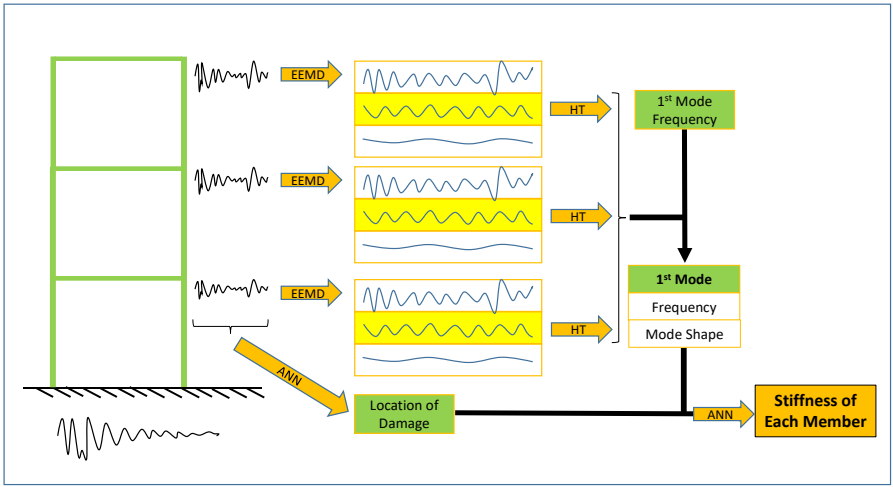

*Figure 8 Schematic view of propsed method*

The first step is recording response of the structure using installed sensors. It should be mentioned that because excitation on the base level of the structure is an earthquake record as well as damping effect on structure the corresponding response is non-stationary. Another point which deserves to be alluded is that the procedure is output-only and order to underline this fact, there are not any installed sensors on the base although it could be. As mentioned earlier the first exquisite step is recording the structural response in each story level. These signals is considered as input data for EEMD and the extracted results are IMFs. Then, HT is applied on IMFs to find instantaneous frequencies and amplitudes. It should be mentioned in order to evaluate the first mode frequency, each story signal is satisfying and the result can be double-checked by the others. Additionally, in most cases, the structural response are more obvious in higher elevation, the last story response, thus, is considered as the main signal for finding the dominant frequency. It is further shown that the result are not significantly sensitive to about pinpointing the first mode dominant frequency.

In the next step, thanks to more uniform IMFs by virtue of EEMD, normalized each story IMF, which contain the first mode frequency, relative to the last one IMF. By doing so, first mode shape is in hand. So far, the first mode parameters including shape and frequency are evaluated.

Then, a neural network is employed in order to predict the response of the each story. This network is a radial basis function network (RBFN) training procedure of which is facile and using it is straightforward nonetheless the most common architecture for a network is the so-called multi-layer perceptron (MLP)–which is a feedforward multi-layer network with hyperbolic tangent activation functions. The duty of this network is to forecast the acceleration of each story using its acceleration in previous moments and other stories in current and previous moments. To put it differently, by using early seconds response, when the excitation is not destructive, or a former excitation which is not strong enough to damage the structure this network is trained and ready to predict the behavior of the structure. As mentioned earlier, any differences between predicted value and measured value signify changes in structural elements on the grounds that RBFN are remarkably powerful to predict acceleration response precisely. This step is extremely worthwhile to locate the damage and facilitate the procedure for the next step.

The last but not the least step is another network so as to measure severity of damage. Although the aforementioned steps are non-model based method, in this step the intact and undamaged structure is needed to be compared with damaged one. Furthermore, this network needs some data to be trained so as to be serviceable. In other words, if the relationship between the first mode shapes and frequency and its corresponding parameters of the associated structure is available, the structural parameters can be identified from the first mode shapes and frequency; however, a formidable challenge is to establish a mathematical model for mapping from the mode shapes and frequency to the structural parameters. Although with known stiffness matrix the eigen values and eigen vectors are simply in hand, and in doing so, mode shapes and frequencies are calculated, it is formidable task to reach a general solution to assess the stiffness matrix by first mode shapes and frequency. Therefore, using such optimization algorithms can be helpful to solve this inverse problem as Genetic Algorithm (GA), Swarm Intelligence (SI), Particle Swarm Optimization (PSO), Multiparticle Swarm Coevolution Optimization (MPSCO), and Improved Multiparticle Swarm Coevolution Optimization (IMPSCO)–which is employed by Jiang et al. [30] to localize and quantify the structural damage and compared with GA–, to name but a few. In this study an ANN is applied on the grounds that the neural network has the alluded ability to approximate arbitrary continuous function and mapping as well, so it stands to the reason that the mathematical model is supplanted by a network which can be considered as an optimization problem with the severity of the damages being its variables.

4. Modeling and Analysis

The first structure is a three-story steel moment-resisting frame, which is depicted in Figure 9, and its specifications are provided in Table 1. The open source finite element program, Opensees is used for nonlinear dynamic analysis. The utilized material is Steel01 which is a bilinear model, and by using 0% strain hardening it is transformed to elastic-perfectly plastic model, which these parameters are summarized in Table 2. This structure has Rayleigh proportional damping with both the first and the third modal damping ratios of 0.05. Scaled Northridge earthquake record (It is multiplied by to in order to push the structure into plastic range) is applied to the structure and the ground motion and the structure response are shown in Figure 10 and Figure 11 respectively.

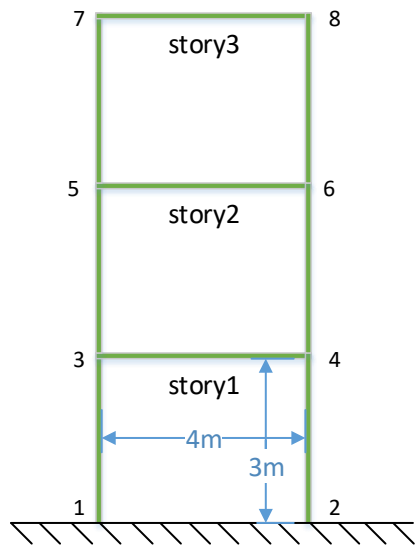

*Figure 9 Overview of the three-story frame*

*Table 1 Specifications of the three-story frame*

| Story | Mass(kg) | Stiffness of Beam Section (MPa) | Stiffness of Column Section (MPa) |
|---|---|---|---|
| 1st | 8000 | 8.4164062 | 17.45226 |
| 2nd | 8000 | 8.4164062 | 17.45226 |
| 3rd | 8000 | 8.4164062 | 17.45226 |

*Table 2 Utilized material*

| Material No. | Fy (MPa) | E Elastic (MPa) | E Plastic (MPa) | $\delta_y$ |
|---|---|---|---|---|
| 1 | 200 | $2 \times 10^5$ | 0 | 0.01 |

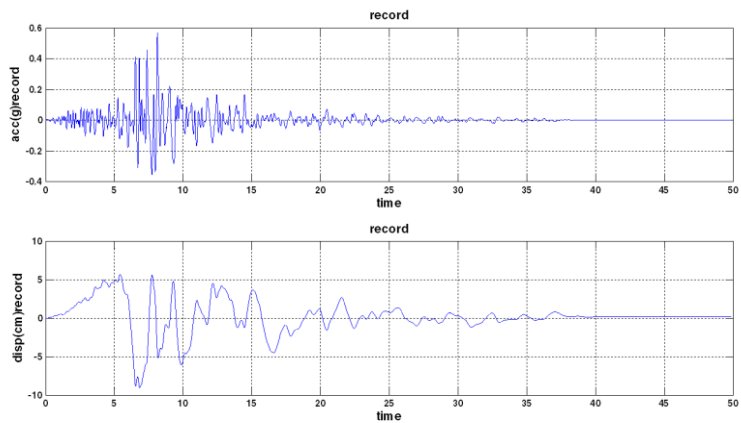

*Figure 10 Scaled Northridge ground motion accelaration and displacement record*

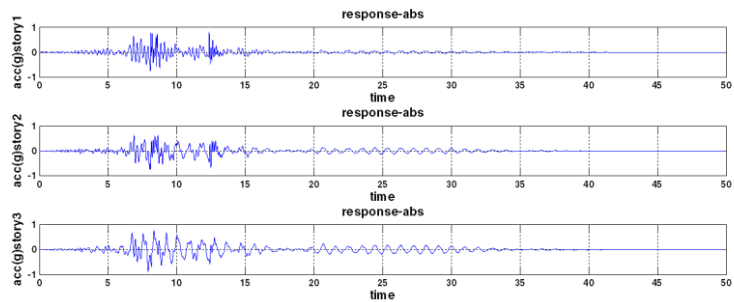

*Figure 11 The structure acceleration response*

Apart from alluded points, in order to simulate damage in column, each return to the elastic range after undergoing plastic behavior reduces column stiffness to 91% of its previous stiffness. In this regard, it should be mentioned that although several both beams and columns are stressed beyond their elastic limit and into the plastic range and tolerate nonlinear behavior during a seismic excitation, simply columns are susceptible to degradation. This changes in structural element stiffness are demonstrate in Figure 12 in which second and third story column is not shown because they remain in elastic range.

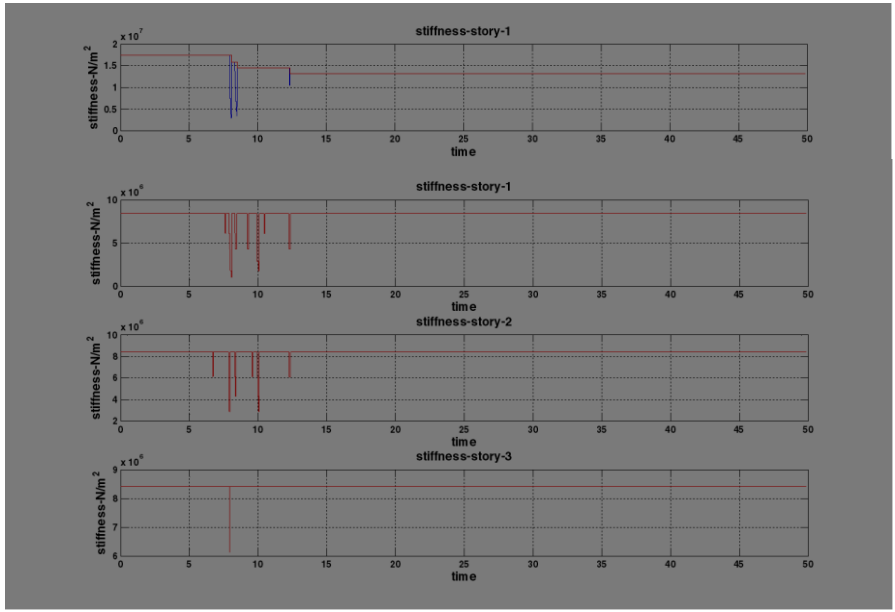

Figure 12 Changes in beams and fist story columns stiffness

In this study, two salient future of EEMD is considered, that is noise level and ensemble number. As the names imply, they determine amplitude of noise according to standard deviation of original data in each step and number of iteration respectively. In spite of the fact that an ensemble number of a few hundred bring about decent results, there is no definite value for noise level [20]. Figure 13 shows IMFs using EMD of the base and story1 to story3 from left to right. In fact, if noise level is 0 and ensemble number is 1 EEMD change into EMD, and displays its corresponding HT. Figure 15 to Figure 24 show IMFs using EMD of the base and story1 to story3. In all these figures the first row is original signal and following rows are it corresponding IMFs. Observing the results accurately, one can find out that though IMFs improve by increasing ensemble number, the 5000 is enough, and values more than this just increase computational cost. Furthermore, almost noise level which is less than 0.5 is not an acceptable remedy, so it is better to use values more than 0.5. Figure 27 shows HT of IMFs of Figure 21. There is a histogram close to every HT which is a new method. The dominant frequency is assessed by these histograms rather than conventional marginal distribution. On account of the fact that not only does the structure undergo plastic deformation, but also because of instantly changing in stiffness matrix, instantaneous frequency of the structure alters. Thus, it is not impossible that HT provides large values which affect marginal distribution. On the contrary, constructing a histogram is based on counting how many values fall into each interval; therefore, effects of inordinate amplitude is squandered in estimating dominant frequency. In Figure 27, apart from the first column which is related to ground motion, IMFs number 7 and 8 exhibit the lowest meaningful dominant frequency which emanates from first mode vibration. Furthermore, Figure 28 presents normalized total amplitude of IMFs number 7 and 8 for each story relative to story 3 for different values of noise level and ensemble number. According to this figure, as mentioned earlier, the noise level at or below 0.5 does not provide decent results. Meanwhile, when the noise level is 1 or 2 the results are closely resemble each other and helpful. Moreover, Figure 28 depicts the mentioned normalized amplitude according to ensemble number. In this regard, when ensemble number is 5000 and 10000, normalized amplitudes in each column are the same.

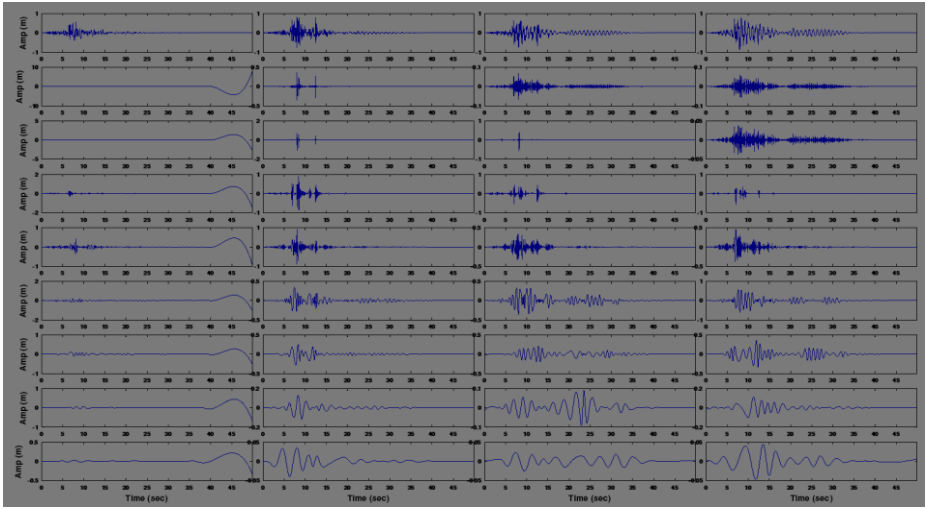

*Figure 13 The ground motion acceleration and structure acceleration response IMFs, noise level=0 ensemble number=1*

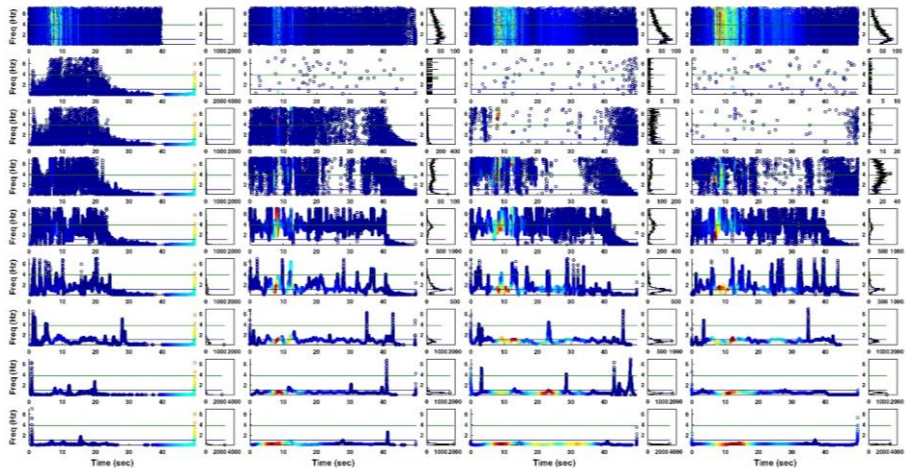

*Figure 14 Hitlbert trasform of EMD*

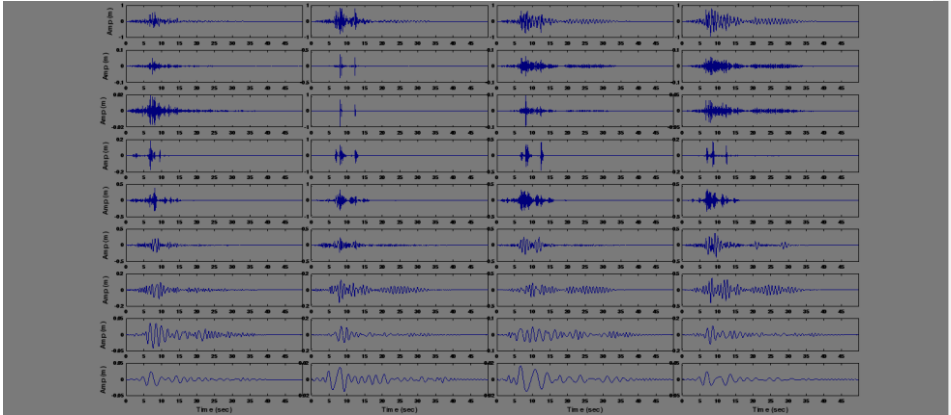

*Figure 15 The ground motion acceleration and structure acceleration response IMFs, noise level=0.1 ensemble number=1000*

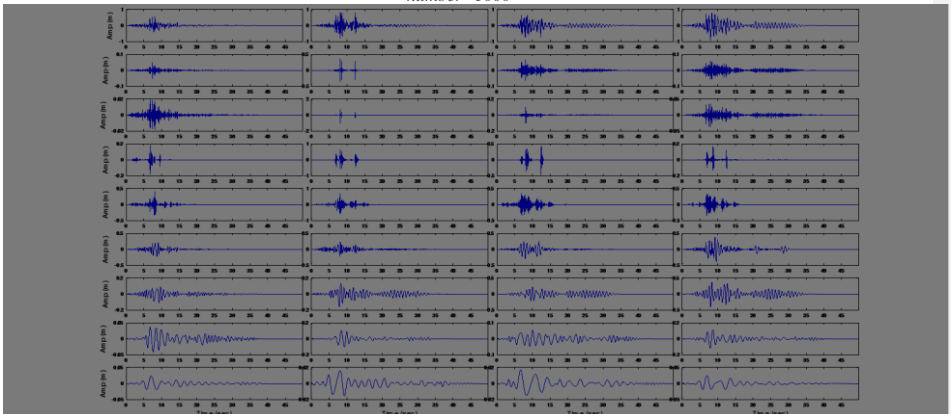

*Figure 16 The ground motion acceleration and structure acceleration response IMFs, noise level=0.1 ensemble number=5000*

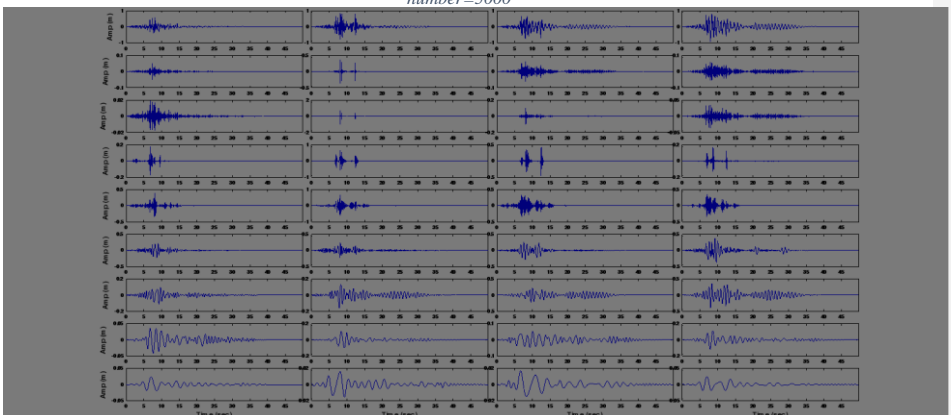

*Figure 17 The ground motion acceleration and structure acceleration response IMFs, noise level=0.1 ensemble number=10000*

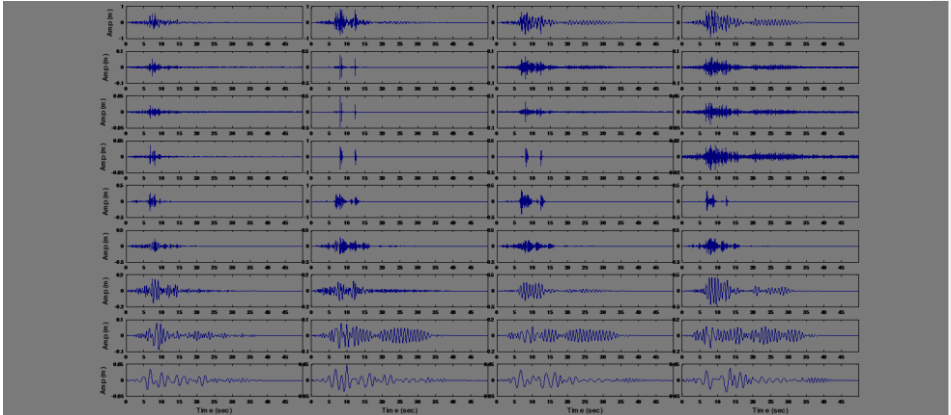
*Figure 18 The ground motion acceleration and structure acceleration response IMFs, noise level=0.5 ensemble number=1000*

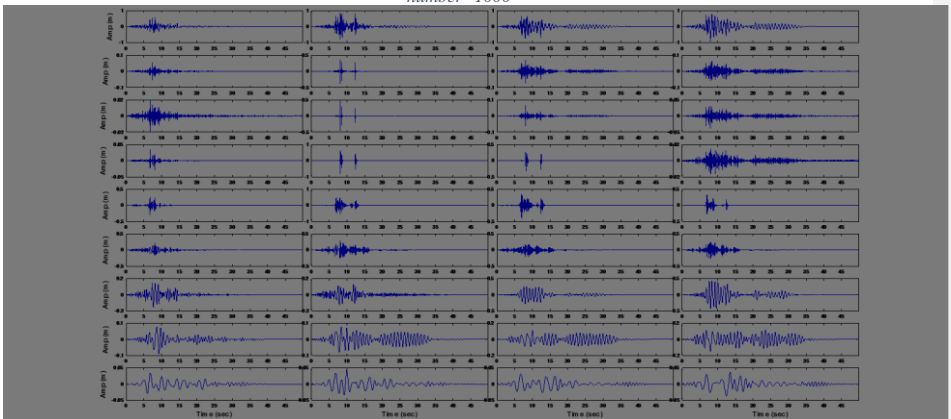
*Figure 19 The ground motion acceleration and structure acceleration response IMFs, noise level=0.5 ensemble number=5000*

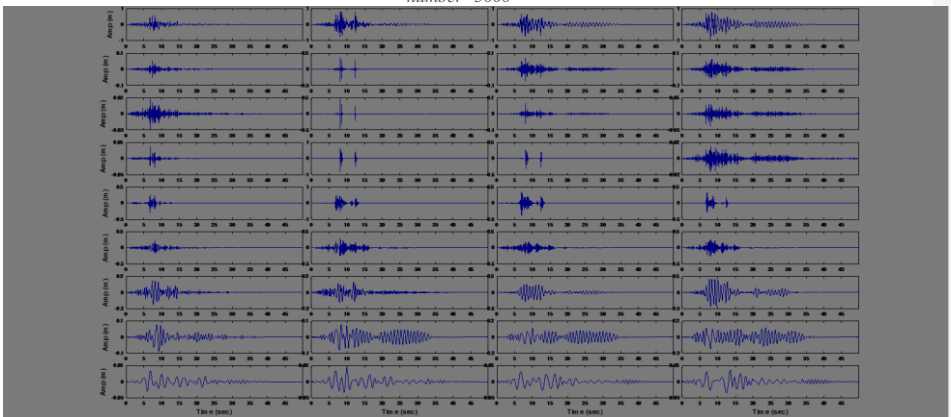
*Figure 20 The ground motion acceleration and structure acceleration response IMFs, noise level=0.5 ensemble number=10000*

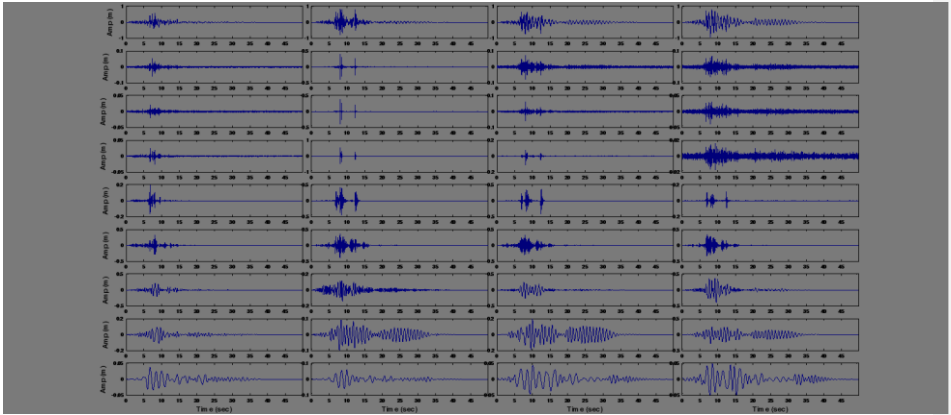
*Figure 21 The ground motion acceleration and structure acceleration response IMFs, noise level=1 ensemble number=1000*

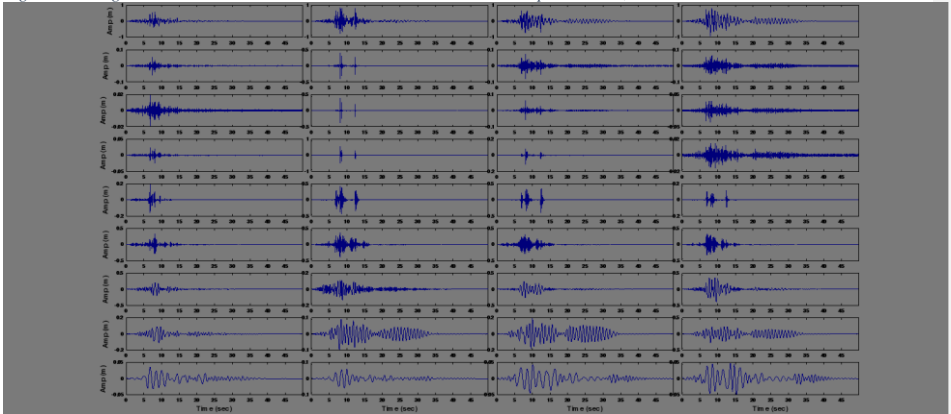
*Figure 22 The ground motion acceleration and structure acceleration response IMFs, noise level=1 ensemble number=5000*

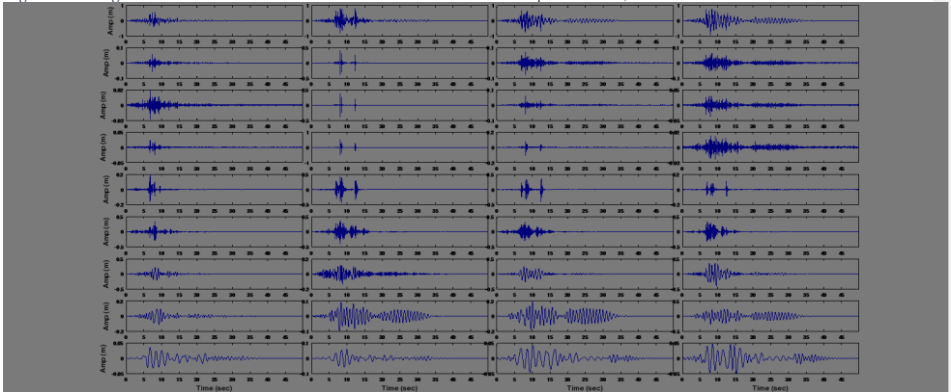
*Figure 23 The ground motion acceleration and structure acceleration response IMFs, noise level=1 ensemble number=10000*

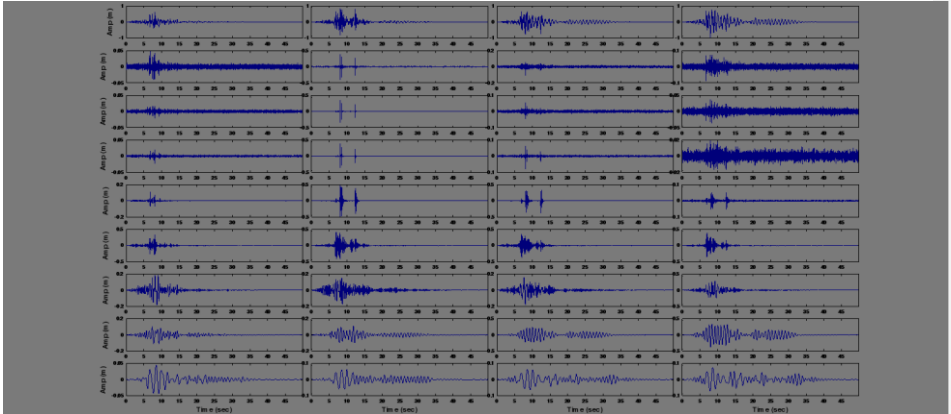
*Figure 24 The ground motion acceleration and structure acceleration response IMFs, noise level=2 ensemble number=1000*

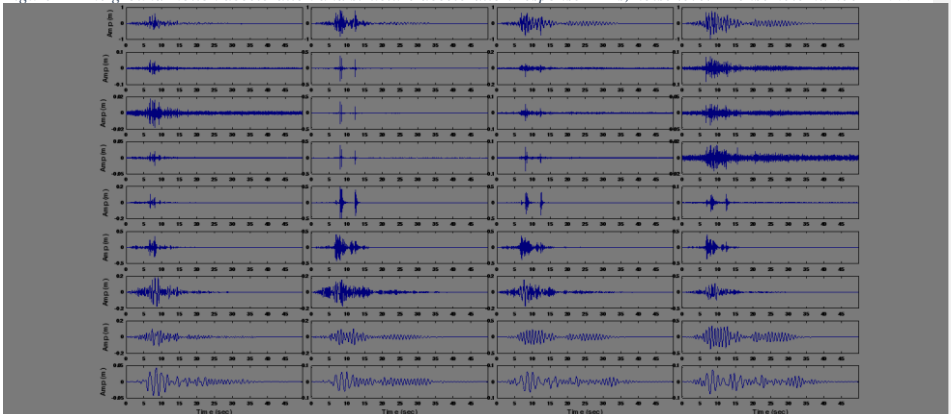
*Figure 25 The ground motion acceleration and structure acceleration response IMFs, noise level=2 ensemble number=5000*

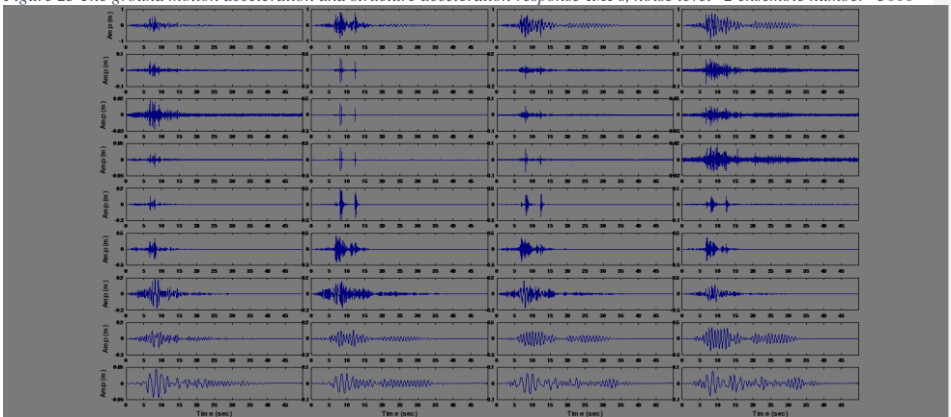
*Figure 26 The ground motion acceleration and structure acceleration response IMFs, noise level=2 ensemble number=10000*

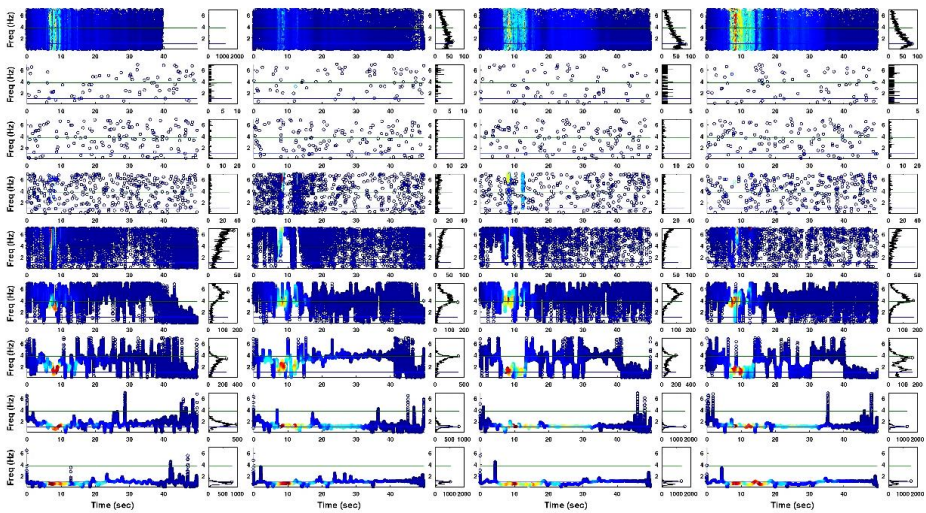

*Figure 27 Hitlbert trasform of EEMD with noisel level=1 and ensemble number=5000*

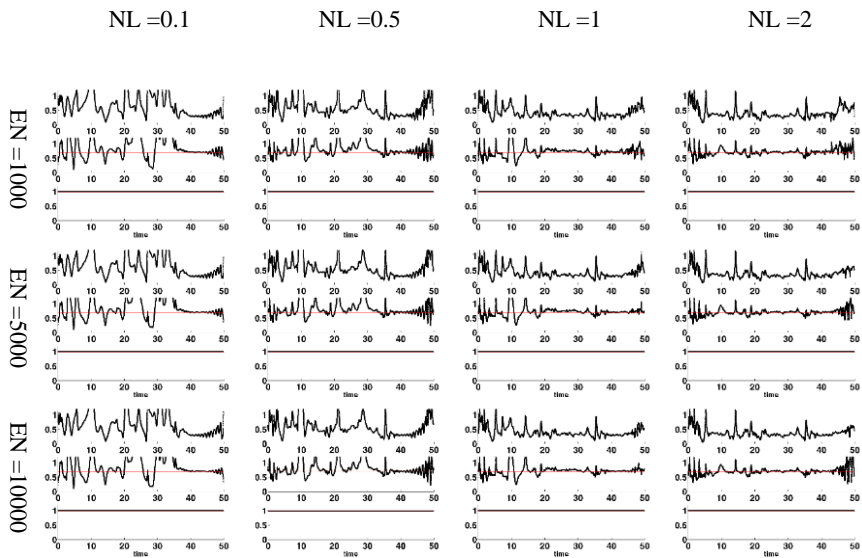

*Figure 28 Normalized total amplitude of IMFs number 7 and 8 for each story relative to story 3*

According to the Figure 28, the longest duration which variance of values in normalized amplitudes reach a stable and minimum level is considered as the normalized mode shape, so when ensemble number is 5000, and noise level is 1 this timespan spanning 6 seconds from 36. Table 3 compares first mode shape and frequency calculated by Finite Element Method (FEM) with proposed method briefly.

*Table 3 Comparsion of FEM results and propsed method results*

|  | Frequency 1st Mode (Hz) | Mode Shapes | | |
| --- | --- | --- | --- | --- |
|  |  | 1st story | 2nd story | 3rd story |
| Calculated by FEM model | 1.0696 | 0.2964 | 0.7037 | 1 |
| Calculated by EEMD & Hilbert | 1.07 | 0.2978 | 0.7055 | 1 |
| Percentage of error | 0.5* | 0.5 | 0.3 | 0 |

*Center of [1.065 1.075] is 1.07. Therefore, maximum error is (1.075-1.0696)/1.0696=0.005

The next step is employing two networks, which discussed before. The schematic architecture of the first network is illustrated in Figure 29. In fact, this network needs a vector with $m \times n + n - 1$ member, where $m$ is the required previous moments, and $n$ is the number of the stories b. Consequently, $m \times n$ and $n-1$ define total number of input data based on previous moments and current moment respectively. In this study $n$ is 3 due to the structure stories and $m$ is 4 notwithstanding the fact that many other values has been tested. This network is trained for each story (for this 3-story frame, 3network is needed and trained. Their architecture are similar and difference is just in input and output data) and predicted acceleration response is shown in Figure 30. Another subtle point which deserves to be mentioned is that in order to predict the acceleration response in story $k$, previous moment acceleration response for story $k-1$, $k$, and $k+1$ as well as current moment acceleration response for story $k-1$ and $k+1$. In other words, a lower and upper story response data is enough, but alluded measures has taken because of lacking sensor in base level of this frame. Therefore, using a sensor in the base level not only improves the performance of the network but also diminishes the input data.

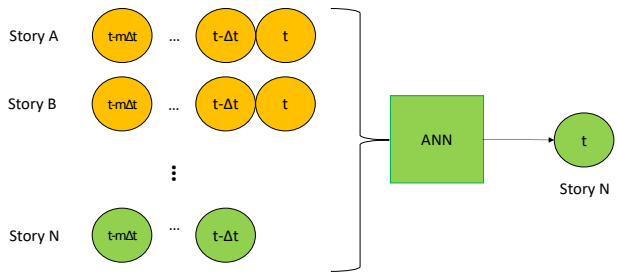

*Figure 29 Overview of emulator artificial neural network*

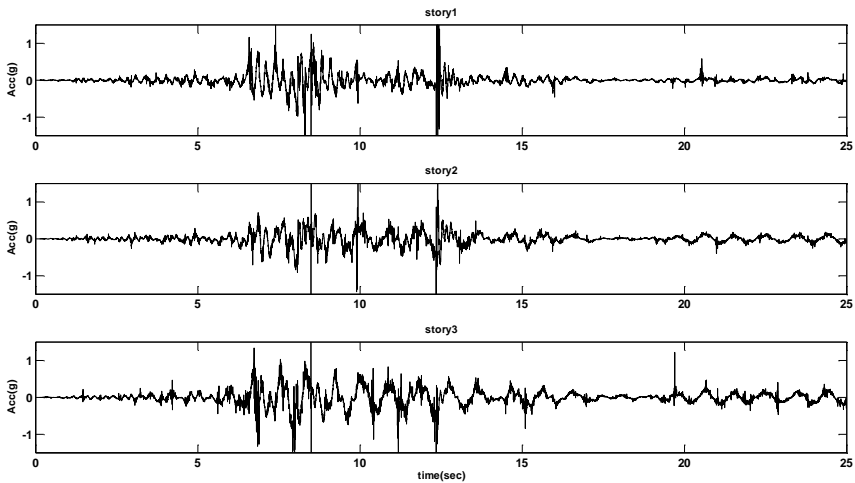

*Figure 30 The predicted accelaration response of the structure*

Figure 31 presents the difference between structure response and neural network prediction, that is, absolute difference, cumulative difference (showing a gentle slope owing to added noise to input and output data), and modified cumulative difference (eliminating the gentle slope) in first row to third respectively. The big jumps in story 1 error reveal changes in structure behavior and the vicinity of two other story errors shows their anomaly in predicted data stem from one common source which is another story, story 1.

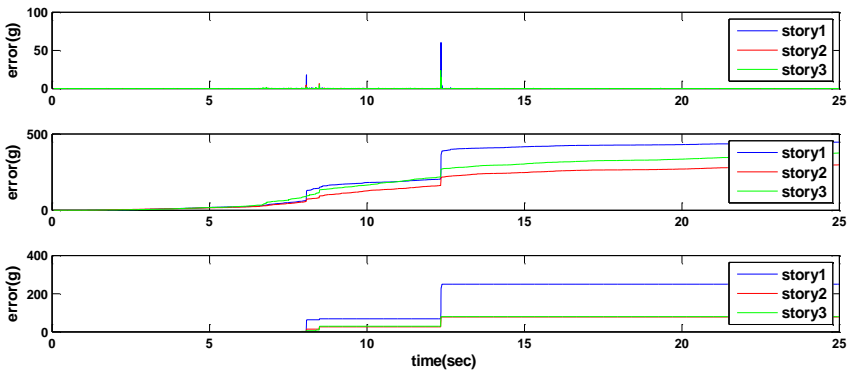

*Figure 31 The difference between structure response and neural network prediction*

Final step, after location of damage and first mode parameters are determined, is the second network, as shown in Figure 32. This network is trained by 100 set of training patterns, which associates with performance of previous network. The summarized results are presented in Table 4 and, in addition, Figure 33 shows sensitivity of the proposed algorithm to first mode dominant frequency. Indeed, because of using histogram rather than marginal distribution, the first mode frequency is the interval center; nevertheless, this figure shed light on the fact that length of beans is not determining factor.

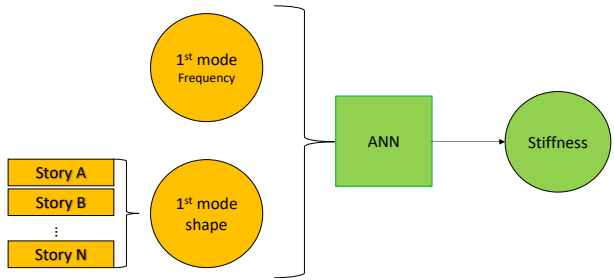

*Figure 32 Overwiew of second artificial neural nerwork*

*Table 4 Comparsion of FEM results and propsed method results*

| Damaged Story | Evaluated Stiffness (MPa) | Calculated by FEM (MPa) | Percentage of error |
|---|---|---|---|
| #1 | 12.70 | 13.15 | 3.4 |

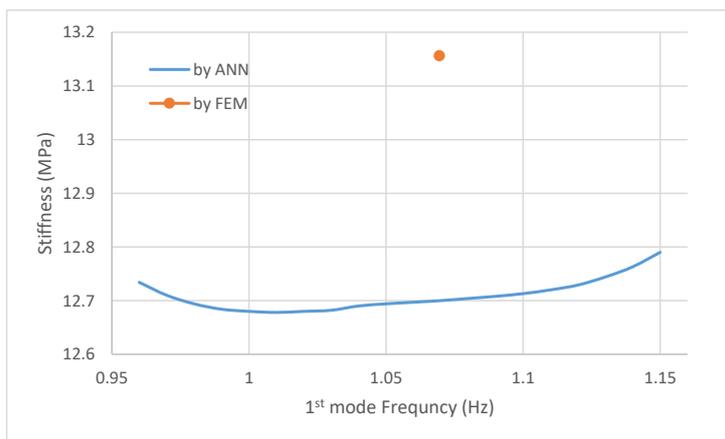

*Figure 33 Sensitivity of the proposed algorithm to first mode dominant frequency*

5. Conclusion

To recapitulate, in this study, EMD is supplanted by EEMD in the process of HT. Even though these methods are closely resemble to each other, EEMD brings more appropriate IMFs which are employed to assess first mode frequency and mode shape. Afterward, an ANN is applied to predict story acceleration based on acceleration of structure during pervious moments. ANN functions precisely. Therefore, any congruency between predicted and measured acceleration provides onset of damage. Then another ANN method is applied to estimate stiffness matrix. Though first mode shape and frequency is calculated in advance, it essentially requires that an inverse problem to be solved in order to find stiffness matrix. This task is done by another ANN. In other words, these two ANN method are exercised to forecast location and measure severity of damage respectively. This algorithm is implemented on two nonlinear moment-resisting steel frame and the results are acceptable. In actual fact, a novel technique based on combination of a time-series method and ANNs which not only overwhelms the limitations of time-frequency methods for damage detection of nonlinear structures but also provides an online monitoring method which is matter of importance in taking preventive measures. Furthermore, that EEMD is implemented rather than EMD affords calculation of first mode shape by normalizing them relative to the last story.